\documentclass{aa}  

\usepackage{subfig}
\usepackage{xspace}
\usepackage{amsmath}
\usepackage{amssymb}
\usepackage{graphicx}
\usepackage{natbib}
\usepackage{lscape}

\usepackage{multirow}

\begin{document}

\newcommand{\um}{\ensuremath{\mu\mathrm{m}}}
\newcommand{\lnir}{\ensuremath{L_{\rm AGN}^{\rm NIR}}}
\newcommand{\lmir}{\ensuremath{L_{\rm AGN}^{\rm MIR}}}
\newcommand{\lx}{\ensuremath{L_{\rm AGN}^{\rm X}}}
\newcommand{\oiv}{\ensuremath{{\rm [O IV]} \lambda 25.89 \um{}}}
\newcommand{\oiii}{\ensuremath{{\rm [O III]} \lambda 5007}}
\newcommand{\ha}{\ensuremath{{\rm H} \alpha}}
\newcommand{\hb}{\ensuremath{{\rm H} \beta}}
\newcommand{\loiv}{\ensuremath{L_{\rm AGN}^{\rm [OIV]}}}
\newcommand{\fnir}{\ensuremath{f_{\rm AGN}^{\rm NIR}}}
\newcommand{\brg}{\ensuremath{{\rm Br}\gamma}}
\newcommand{\nh}{\ensuremath{N_{\rm H}}}
\newcommand{\av}{\ensuremath{A_{\rm V}}}
\newcommand{\nhav}{\ensuremath{N_{\rm H}/A_{\rm V}}}
\newcommand{\nhavgal}{\ensuremath{\left(N_{\rm H}/A_{\rm V}\right)_{\rm galactic}}}

\newcommand{\prepclaudio}{Ricci et al., in prep}
\newcommand{\prepmike}{Koss et al., in prep}
\newcommand{\prephagai}{Netzer et al., in prep}
\newcommand{\prepallan}{Schnorr-M\"uller et al., in prep}

\title{On the relation of optical obscuration and X-ray absorption in Seyfert galaxies}

\author{L. Burtscher \inst{1}
	\and
	R. I. Davies \inst{1}
	\and
	J. Graci\'{a}-Carpio \inst{1}
	\and
	M. J. Koss \inst{2}
	\and
	M.-Y. Lin \inst{1}
	\and
	D. Lutz \inst{1}
	\and
	P. Nandra \inst{1}
	\and
	H. Netzer \inst{3}
	\and
	G. Orban de Xivry \inst{1}
	\and
	C. Ricci \inst{4}
	\and
	D. J. Rosario \inst{1}
	\and
	S. Veilleux \inst{5,6}
	\and
	A. Contursi \inst{1}
	\and
	R. Genzel \inst{1}
	\and
	A. Schnorr-M\"uller \inst{1}
	\and
	A. Sternberg \inst{7}
	\and
	E. Sturm \inst{1}
	\and
	L. J. Tacconi \inst{1}
}

\institute{Max-Planck-Institut f\"ur extraterrestrische Physik,
		Postfach 1312, Gie\ss enbachstr., 85741 Garching, Germany, 
		\email{burtscher@mpe.mpg.de}
	\and
		Institute for Astronomy, Department of Physics, ETH Z\"urich, Wolfgang-Pauli-Strasse 27, CH-8093 Z\"urich, Switzerland
	\and
		School of Physics and Astronomy, Tel Aviv University, Tel Aviv 69978, Israel
	\and
		Instituto de Astrof\'isica, Facultad de F\'isica, Pontificia Universidad Cat\'olica de Chile, 306, Santiago 22, Chile
	\and
		Department of Astronomy, University of Maryland, College Park, MD 20742, USA
	\and
		Joint Space-Science Institute, University of Maryland, College Park, MD 20742, USA
	\and
		Raymond and Beverly Sackler School of Physics \& Astronomy, Tel Aviv University, Ramat Aviv 69978, Israel
}

\date{Draft version, \today{}}                                           

\abstract{The optical classification of a Seyfert galaxy and whether it is considered X-ray absorbed are often used interchangeably. But there are many borderline cases and also numerous examples where the optical and X-ray classifications appear to be in conflict. In this article we re-visit the relation between optical obscuration and X-ray absorption in AGNs. We make use of our ``dust color'' method (Burtscher et al. 2015) to derive the optical obscuration \av{} and consistently estimated X-ray absorbing columns using 0.3--150 keV spectral energy distributions. We also take into account the variable nature of the neutral gas column \nh{} and derive the Seyfert sub-classes of all our objects in a consistent way.

We show in a sample of 25 local, hard-X-ray detected Seyfert galaxies ($\log L_{\rm X} / {\rm(erg/s)} \approx 41.5 - 43.5$) that there can actually be a good agreement between optical and X-ray classification. If Seyfert types 1.8 and 1.9 are considered unobscured, the threshold between X-ray unabsorbed and absorbed should be chosen at a column $\nh{} = 10^{22.3}$ cm$^{-2}$ to be consistent with the optical classification.

We find that \nh{} is related to \av{} and that the \nhav{} ratio is approximately Galactic or higher in all sources, as indicated previously. But in several objects we also see that deviations from the Galactic ratio are only due to a variable X-ray column, showing that (1) deviations from the Galactic \nhav{} can simply be explained by dust-free neutral gas within the broad line region in some sources, that (2) the dust properties in AGNs can be similar to Galactic dust and that (3) the dust color method is a robust way to estimate the optical extinction towards the sublimation radius in all but the most obscured AGNs.}


\maketitle

\section{Introduction}

Seyfert galaxies are commonly classified by the presence or absence of broad permitted emission lines in the optical, though finer classification schemes exist to differentiate between the various types of broad-line galaxies. A complementary classification scheme is afforded by space-based X-ray observations which allow us to determine the equivalent neutral hydrogen absorbing column \nh{} towards the compact X-ray emitting corona around the central black hole. The result of these phenomenological classifications is a great apparent diversity of active galactic nuclei (AGNs).

The most successful approach to understand this menagerie of AGN types is unification by orientation \citep{rees1966}. Polarized, i.e. scattered, broad lines are probably the strongest indication for a ``torus'', i.e. toroidal obscuration by dust \citep{antonucci1985} blocking our line of sight towards the Broad Line Region (BLR) at some inclinations. Strong support for a confining structure on small scales also comes from observations of well-defined ionization cones \citep[e.g.][]{schmitt2003}. For a comprehensive overview of the concept of the torus we refer to the recent review by \citet{netzer2015}.

Plenty of observational tests have challenged the ``unified model''. For example, claims for obscured (``type 2'') AGNs without hidden BLRs have been made \citep{tran2001}, but recent investigations have shown that most of these objects actually {\em do} show broad lines in polarized light if sensitivity is carefully taken into account (Ramos Almeida et al., submitted). Objects that intrinsically lack a BLR (dubbed ``true type 2'') are rare \citep{panessa2002} and most of them turn out to be either ordinary ``type 2s'' (with a substantial column of neutral gas) or faint ``type 1s'' if examined carefully \citep{shi2010}.

However, there are clearly exceptions to the rule such as type 1 AGNs in edge-on configurations \citep[e.g. NGC~3783, ][]{muellersanchez2011}. This argues in favor of statistical unification where an AGN is more likely to be observed as a broad-line (type 1) AGN if its torus has a larger opening angle or a smaller number of clouds as shown by SED-fitting with clumpy torus models \citep{ramosalmeida2011,elitzur2012}.

It is also a matter of debate where exactly the obscuring structure is located. While infrared continuum reverberation mapping and interferometry have shown that the hot and warm dust are located on (sub-)parsec scales \citep[e.g.][and references therein]{burtscher2013}, geometrically thick gas structures are also found on the slightly larger ($\sim$ 30 pc) scales of nuclear star clusters \citep{hicks2009} and there is also evidence in a few Compton-thick galaxies that a substantial amount of obscuration might be contributed by the host galaxy since some of the deepest silicate absorption features are seen in edge-on galaxies \citep{goulding2012}.

Much of the uncertainty arises due to different tracers of the obscuration/absorption down to the nucleus and may lead to inconsistent classification across the electromagnetic spectrum \citep[e.g.][]{merloni2014}. While some of this inconsistency may be caused by dust-free, neutral absorption in the torus \citep{davies2015}, one has to keep in mind that the different tracers are susceptible to various issues, e.g. the effects of host galaxy dilution and contamination, or the details of X-ray spectral modeling.

In this work, we use a new measure of the NIR--MIR extinction \citep{burtscher2015}, which is more sensitive to high dust columns than purely optical methods. Also, we use the best available X-ray spectral modeling results for a well-defined sample of local AGNs. We fold in a consistent classification of Seyfert types and -- most importantly -- take into account the temporal variability of the X-ray absorbing column. With this, we explore the relationships between the X-ray and optical obscuration in a more rigorous way than done before. We compare the columns of gas and dust that the different tracers of obscuration imply and discuss what this means for the concept of the torus and its structure.

The paper is organized as follows: We briefly introduce the sample and method (Section \ref{sec:sample}) and then study the relation between the optically-classified Seyfert type and X-ray absorption (Section \ref{sec:NH_subtype}) as well as the relation between optical obscuration and the X-ray absorbing column and discuss its implications (Section \ref{sec:AVNH}).


\section{Sample and Analysis}
\label{sec:sample}

\begin{table*}
\caption{\label{tab:sample}Sample properties}
\begin{tabular}{llllllll}
\hline
\hline
ID & \fnir{} & $T$ & $A_{\rm V}$& $\log \nh{}$  & $\log \nh{}$ & Seyfert & $\oiii{}/\hb{}$ \\
&&&&BAT&variable&type&\\
 & & [K] & [mag] & \multicolumn{2}{c}{[cm$^{-2}$]} && \\
\hline
Circinus~galaxy & 0.77 & 691 & 27.2 & 24.4 &---& 1h$^a$ & --- \\
ESO~137-G034 & $<0.1$ & --- & --- & 24.3 &---& 2$^b$ & --- \\
ESO~548-G081 & 0.87 & 1236 & 1.6 & $<$ 20.0 &---& 1.0 & 5.5$^c$ \\
IC~5063 & 0.56 & 947 & 11.3 & 23.6 &$23.3^1-23.4^1$& 1h$^d$ & --- \\
MCG~-05-23-16 & 0.86 & 1207 & 2.3 & 22.2 &$22.1^1 - 22.7^1$& 1.9 & 0.07$^e$ \\
NGC~1052 & 0.24 & 1239 & 1.5 & 23.0 &---& 1h$^f$ & --- \\
NGC~1068 & 0.92 & 723 & 24.6 & 25.0 &---& 1h$^g$ & --- \\
NGC~1365 & 0.94 & 1282 & 0.4 & 22.2 &$22.0^2 - 25.0^1$& 1.8 & 0.08$^e$ \\
NGC~1566 & 0.57 & 1479 & 0.0 & $<$ 20.1 &$<20.1^4 - 21.9^{12}$& 1.5 & 0.91$^h$ \\
NGC~2110 & 0.81 & 967 & 10.5 & 23.0 &$22.5^3 - 22.9^4$& 1h$^s$ & --- \\
NGC~2992 & 0.61 & 1111 & 5.1 & 21.7 &$21.6^1 - 22.2^1$& 1.8 & 0.09$^e$ \\
NGC~3081 & 0.37 & 977 & 10.0 & 23.9 &$23.7^1 - 23.8^1$& 1h$^i$ & --- \\
NGC~3227 & 0.67 & 1543 & 0.0 & 21.0 &$21.0^4 - 23.4^5$& 1.5 & 0.37$^j$ \\
NGC~3281 & 0.83 & 732 & 23.9 & 23.9 &---& 2$^k$ & --- \\
NGC~3783 & 0.99 & 1244 & 1.3 & 20.5 &$20.5^4 - 23.0^6$& 1.5 & 1.07$^e$ \\
NGC~4051 & 0.64 & 1305 & 0.0 &$<$  19.6 &---& 1.5 & 1.1$^l$ \\
NGC~4388 & 0.78 & 887 & 14.2 & 23.5 &$23.3^7 - 23.8^7$& 1h$^m$ & --- \\
NGC~4593 & 0.88 & 1292 & 0.2 & $<$ 19.2 &---& 1.5 & 1.85$^e$ \\
NGC~5128 & 0.82 & 796 & 19.4 & 23.1 &$22.8^8 - 23.2^8$& 2$^n$ & --- \\
NGC~5506 & 0.99 & 1223 & 1.9 & 22.4 &$22.3^1 - 22.7^1$& 1i$^o$ & --- \\
NGC~5643 & $<0.1$ & --- & --- & 25.4 &---& 2$^k$ & --- \\
NGC~5728 & $<0.1$ & --- & --- & 24.2 &---& 2$^k$ & --- \\
NGC~6300 & 0.53 & 677 & 28.4 & 23.3 &$23.3^{1,9} - 25.0^{10}$& 2$^k$ & --- \\
NGC~6814 & 0.58 & 1500$^{\star}$ & 0.0 & 21.0 &---& 1.5 & 0.55$^h$ \\
NGC~7130 & $<0.1$ & --- & --- & 24.0 &---& 1.9$^p$ & --- \\
NGC~7172 & 0.7 & 822 & 17.8 & 22.9 &$22.9^1 - 23.0^1$& 1i$^q$ & --- \\
NGC~7213 & 0.81 & 1181 & 3.0 & $<$ 20.4 &---& 1.5 & 0.79$^h$ \\
NGC~7469 & 0.88 & 1355 & 0.0 & 20.5 &---& 1.0 & 6.5$^r$ \\
NGC~7582 & 0.89 & 1082 & 6.1 & 24.2 &$22.7^1 - 24.1^{11}$& 1i$^s$ & --- \\
\hline
\end{tabular}
\\
{\bf Notes.} \fnir{}: fraction of non-stellar flux in the K band within an aperture of 1\arcsec{} (for most sources), $T$: temperature of a black body from a fit to the K band spectrum (for errors on \fnir{} and $T$, see B15), $A_{\rm V}$: extinction towards the hot dust (from B15; error is estimated to be 3 mag, see Sec.~\ref{sec:sample}), \nh{}: X-ray absorbing column from \prepclaudio{} (formal errors are $<$ 0.2 dex), \nh{} variable: full range of other published \nh{} values for these sources. {\em Seyfert type}: for broad line galaxies (1.0/1.2/1.5) based on a the \hb/\oiii{} ratio (see reference there) and converted to a Seyfert type according to the \citet{winkler1992} scheme; Seyfert 1.8/1.9: \citet{osterbrock1981}, 1i: broad hydrogen recombination line(s) detected at infrared wavelengths, 1h: broad hydrogen recombination line(s) detected in polarized optical light, 2: no broad hydrogen recombination line detected. $^{\star}$ New observations with adaptive optics for NGC~6814 show much hotter dust in this source than in B15. The origin of this discrepancy is unknown and might be related to intrinsic variability. Here we adopt an intermediate value, effectively reducing the adopted $A_{\rm V}$ from 7.2 to 0.\\
{\bf References.} {\em X-ray variability:} $^1$: \citet{risaliti2002}, $^2$: \citet{walton2014}, $^3$: \citet{rivers2014}, $^4$: Ricci et al. (in prep.), $^5$: \citet{lamer2003}, $^6$: \citet{markowitz2014}, $^7$: \citet{fedorova2011}, $^8$: \citet{beckmann2011}, $^{9}$: \citet{guainazzi2002}, $^{10}$: \citet{leighly1999}, $^{11}$: \citet{bianchi2009}, $^{12}$: \citet{kawamuro2013}\\
{\em Seyfert types and line ratios:} $^a$: \citet{oliva1998}, $^b$: \citet{veroncetty2010}, $^c$: \prepmike{}, $^d$: \citet{lumsden2004}, $^e$: \prepallan{}, $^f$: \citet{barth1999}, $^g$: \citet{antonucci1985}, $^h$: \citet{winkler1992}, $^i$: \citet{moran2000}, $^j$: \citet{smith2004}, $^k$: \citet{philips1983a}, $^l$: \citet{grupe2004a}, $^m$: \citet{shields1996}, $^n$: \citet{tadhunter1993}, $^o$: \citet{nagar2002}, $^p$: \citet{storchibergmann1990}, $^q$: \citet{smajic2012}, $^r$: \citet{kim1995}, $^s$: \citet{reunanen2003}
\end{table*}


Our selection is based on the sample presented in \citet[][hereafter: B15]{burtscher2015}. There we used essentially all local ($D \lesssim 60$ Mpc) AGNs observed with VLT/SINFONI integral field spectroscopy in the near-IR K band. Now, we additionally require the source to be detected in the BAT 70-month survey \citep{baumgartner2013} in order to have consistent \nh{} estimates using 0.3--150 keV spectral energy distributions (\prepclaudio{})\footnote{The baseline model that was used for the spectral fitting (in the 0.3-150 keV band) included an absorbed cutoff power-law continuum plus a reflection component (\textsc{pexrav} in XSPEC). For unobscured sources ($\log \nh{}/{\rm cm}^2 \lesssim 22$) we added, if statistically needed, a blackbody component (for the soft excess), a cross-calibration constant (for possible variability between the non simultaneous soft X-ray and BAT observations) and an iron line (or other emission lines in the Fe region). Partially covering ionized absorption was taken into account in the spectral analysis using the \textsc{zxipcf} model in XSPEC. For obscured sources we added a scattered component, a collisional plasma, emission lines and a cross-calibration constant. The \nh{} values quoted here refer to the cold absorber only, since we believe that the warm absorber is rather related to the outflowing component and propably located on scales larger than the BLR/torus absorption.}. We include three additional BAT-detected AGNs (ESO~137-G034, NGC~5728 and NGC~7213) which have new VLT/SINFONI data from our ongoing observing programme \citep{davies2015}. The sample and some relevant properties are listed in Tab.~\ref{tab:sample}.

For these sources we derive the fraction of near-infrared non-stellar light \fnir{} using the level of dilution of the stellar CO(2,0) absorption profile at 2.29\um{} as a proxy for the AGN light. Here, we summarize briefly the method to derive the extinction and refer the reader to B15 for more details.

\begin{figure}
   \includegraphics[trim=0cm 0cm 0cm 0cm, width=\columnwidth]{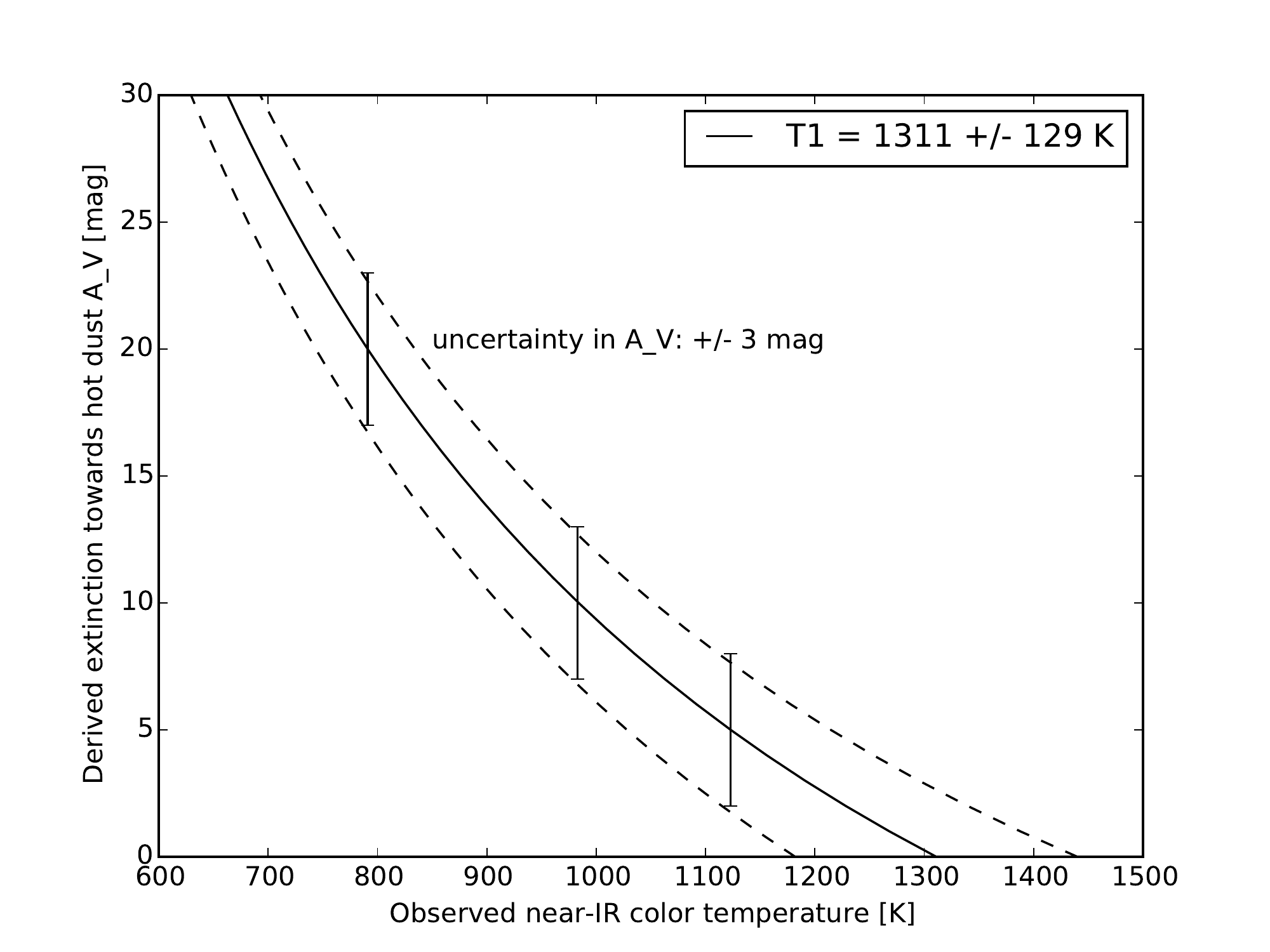}
   \caption{\label{fig:AVT}Derived extinction towards the hot dust \av{} vs. observed near-IR color temperature. The intrinsic temperature $T_1$ is set by the observed color temperature of presumably unobscured sources (Seyfert 1.0 -- 1.5) and the spread in these observed temperatures translates into an uncertainty of $A_V$ of about 3 magnitudes as indicated for several observed color temperatures. For details of the fit of the near-IR color temperature we refer to B15, specifically Fig.~6 there.}
\end{figure}

The temperature is estimated from a fit to the K band spectrum using a stellar template and a blackbody. To derive the optical extinction to the hot dust we use a simple torus model involving a hot ($T \approx 1300$ K) and a warm ($T$ = 300 K) component. By assuming that the intrinsic temperature of the hot dust is roughly a constant, set by the sublimation of dust rather than by the chances of matter distribution, we can explain cooler temperatures and redder near-to-mid-infrared colors ($K-N$) by an absorber with extinction \av{}. The derived extinction as a function of the observed color temperature in the near-IR is shown in Fig.~\ref{fig:AVT}. Since the absorber is likely the parsec-scale ``torus'' and therefore $\approx 4 - 20 \times$ larger than the sublimation radius \citep{burtscher2013}, screen extinction appears to be a suitable absorbing geometry. This is also valid for a clumpy torus where the near-IR emission is still expected to be significantly more compact than the mid-IR emission \citep[e.g. Fig.~6 in][]{schartmann2008}.

For this article we use the average observed color temperature of optical broad-line AGNs (types 1.0--1.9) in this sample as the intrinsic temperature of the hot dust. It is $T_1 = 1311 \pm 129$ K. This scatter in observed intrinsic color temperatures translates to an uncertainty in \av{} of about 3 mag (see Fig.~\ref{fig:AVT}) and constitutes the dominant factor in the uncertainty of \av{}.

Four of the sources in this selection do not show an AGN continuum in the SINFONI data, despite being detected by {\em Swift}/BAT in the very hard X-ray band 14--195 keV. For ESO~137-G034\footnote{The CO map in ESO~137-G034 is complex so that no estimate for the AGN fraction can be given using this method. A spectral fit, however, does not show any indication for a hot near-IR continuum.}, NGC~5643, NGC~5728 and NGC~7130 we can therefore not derive an optical extinction using our method. All of these sources have Compton-thick ($\log \nh{}/{\rm cm}^2 > 24.2$) neutral absorbers, suggesting that the hot dust may be obscured even at infrared wavelengths.

Compared to other methods of estimating the extinction towards the central engine of AGNs, our method has the advantage of being rather simpler: We only require a measurement of the K band continuum at high spatial resolution such as can be afforded with about 10 minutes of VLT/SINFONI time, for example. In order to apply the more widely used method using broad hydrogen recombination lines, on the other hand, {\em multiple} lines must be observed {\em simultaneously} in order to constrain both the dust reddening and the ionization properties in the BLR since ``case B'' is not applicable there \citep[e.g.][]{lamura2007,dong2008}. Such data rarely exist and so spectra are either stitched together from multiple epochs of observations (with all the caveats regarding variability) or case B has to be assumed for the intrinsic line ratios which effectively leads to an upper limit of the actual obscuration \citep[e.g.][]{maiolino2001}.


\section{Seyfert sub-type and X-ray absorption}
\label{sec:NH_subtype}

\begin{figure}
   \includegraphics[trim=0cm 0cm 0cm 0cm, width=\columnwidth]{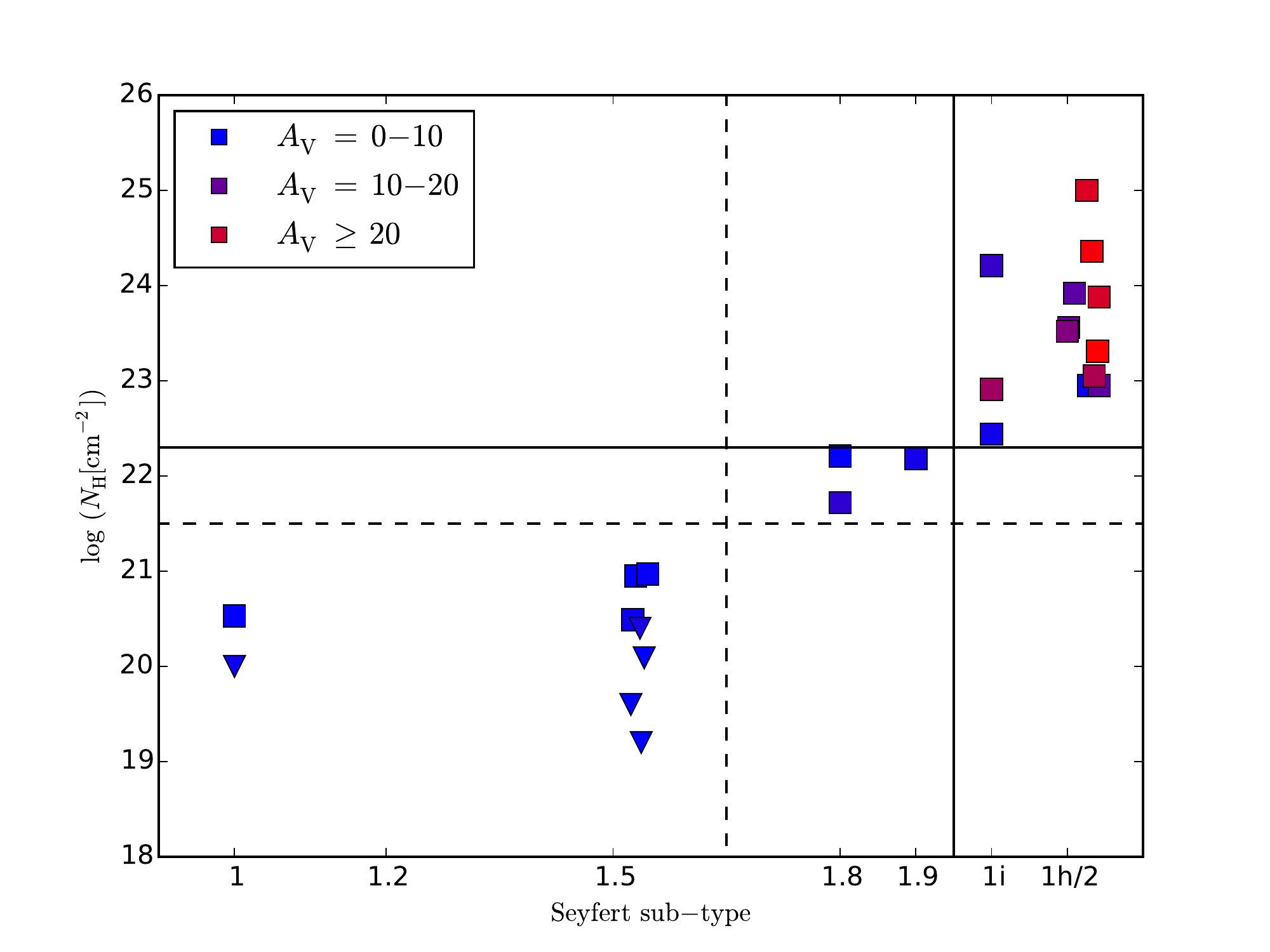}
   \caption{\label{fig:subtype}X-ray absorbing column \nh{} versus Seyfert sub-type. The symbols for sub-types 1.5 and 2 have been randomly shifted by a small amount for better readability. The symbol color indicates the optical obscuration to the hot dust. Marked with solid (dashed) lines are two possible, consistent, dividing lines between optically unobscured/obscured and X-ray unabsorbed/absorbed.}
\end{figure}

In order to investigate the relation between Seyfert sub-type and X-ray classification, we need to choose a classification scheme. Perhaps the most physical scheme is the one by \citet{osterbrock1977} which sub-divides the population into Seyfert 1.0, 1.2 and 1.5 according to the prominence of the broad compared to the narrow component of \hb{}. Since isolation of the broad component of \hb{} is sometimes difficult, however, alternative schemes use the total \hb{} flux and compare it to the flux in the \oiii{} line \citep{whittle1992,winkler1992}. This flux ratio also probes the dominance of direct AGN emission (broad \hb) over ``isotropic'' AGN emission (\oiii) from the Narrow Line Region, but both lines can be contaminated by star formation. It is expected that these contaminations cancel out to some degree which may explain the popularity of this indicator to classify Seyfert galaxies. For this work, we use the \citet{winkler1992} scheme\footnote{This scheme is also employed by \citet{veroncetty2010} on which much of the SIMBAD and NED classifications are based}.

Additionally, the classifications 1.8 and 1.9 are given if $\oiii{}/\hb{} > 3$, but weak broad lines -- \ha{} and \hb{} or only \ha{}, respectively -- are seen \citep{osterbrock1981}. We use the type ``1i'' classification for AGNs that show broad hydrogen recombination lines at infrared wavelengths and finally there are also Seyfert galaxies of type ``1h'' where these lines are only seen in polarized optical light. The resulting classifications are given in Tab.~\ref{tab:sample}.

In Fig.~\ref{fig:subtype} we compare these classifications with the X-ray absorbing column \nh{} (\prepclaudio{}) and find a correlation in the expected sense. It is evident that there is a clear separation between broad-line and narrow-line AGNs in X-ray absorbing column with intermediate-type Seyfert galaxies (1.8/1.9) in between. Despite their low number, this is consistent with expectations and with similar studies in the literature \citep{risaliti1999}.

It is also noteworthy that we do not find broad-line sources with high column density ($\log \nh{}/{\rm cm}^2 \gtrsim 22.3$), such as Mrk~231, BAL~QSOs or the objects reported by \citet{wilkes2002}. The reason for this is probably that such objects are very rare. In optical spectra of all $\approx$ 570 AGNs in the BAT-70 month sample, \prepmike{} find only 9 optically unobscured AGNs with X-ray columns $\log \nh{}/{\rm cm}^2 > 22.3$.

In conclusion, we find a consistent classification between optical and X-ray bands for our sample simply by choosing appropriate boundaries: If $\log \nh{}/{\rm cm}^2 > 21.5$ (dashed lines in Fig.~\ref{fig:subtype}) is chosen for X-ray absorbed AGNs \citep[as in e.g.][]{merloni2014}, the corresponding boundary for ``optically obscured'' should be Seyfert $\geq$ 1.8. On the other hand, if Seyfert 1.8 and 1.9 are to be considered unobscured objects, then an absorbing column $\log \nh{}/{\rm cm}^2 > 22.3$ (indicated by solid lines) should be used to classify objects as X-ray absorbed.

\section{Optical obscuration and X-ray absorption}
\label{sec:AVNH}

\begin{figure*}
\sidecaption
   \includegraphics[trim=0cm 0cm 0cm 0cm, width=12cm]{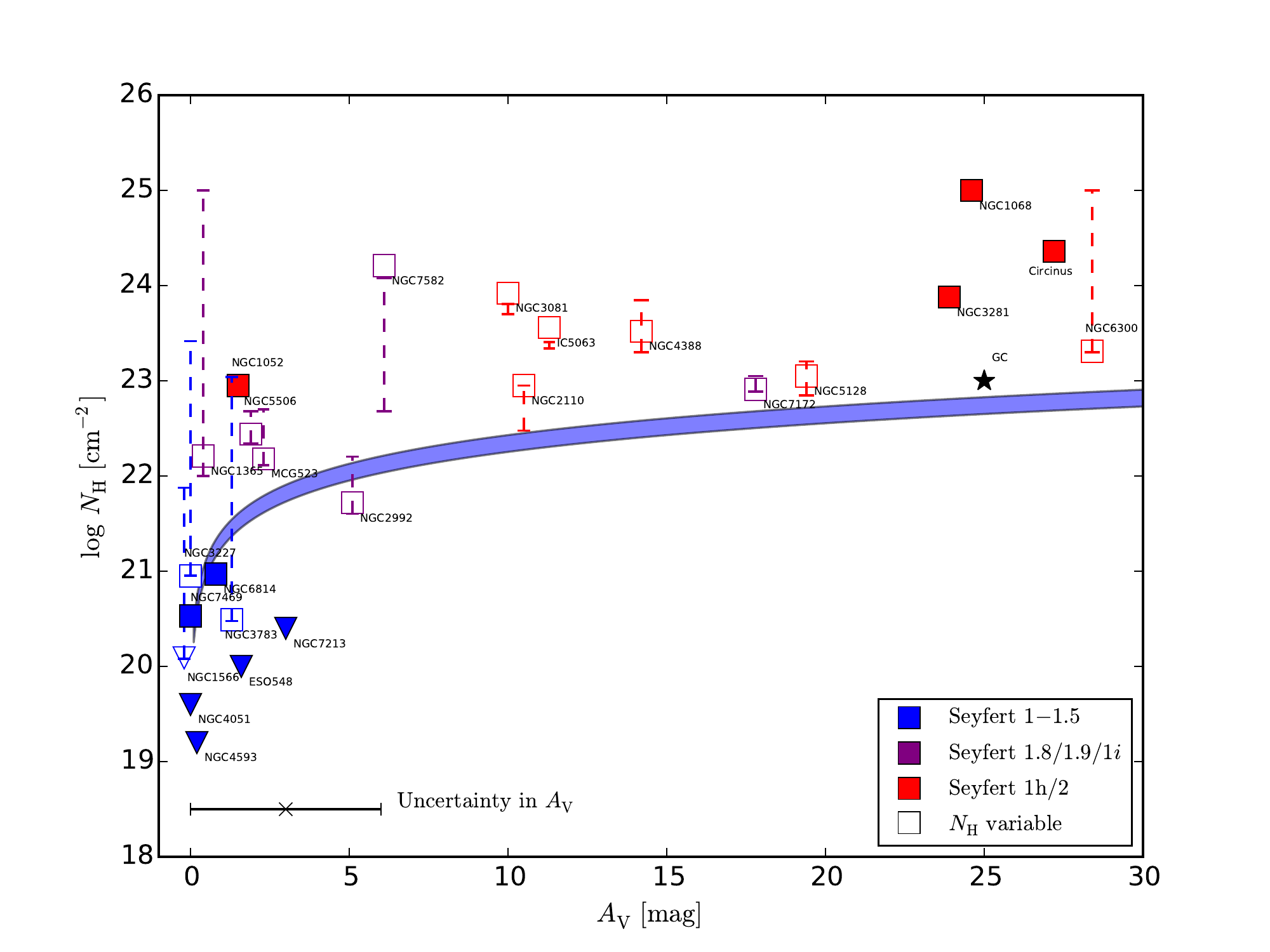}
   \caption{\label{fig:AVNH}\nh{} vs. \av{} with Seyfert sub-type marked by color. The blue band gives the Galactic standard ratio (see text for details). Open boxes indicate sources with variable \nh{} as listed in Tab.~\ref{tab:sample}. The Galactic Center is marked with an asterisk for comparison. Some sources with \av{} = 0 have been slightly offset for clarity.}
\end{figure*}

We can also directly look at the relation between X-ray absorbing column and the optical extinction. Fig.~\ref{fig:AVNH} summarizes our analysis, and shows a number of interesting features.

First, we find a general trend of increasing \nh{} with increasing \av{}: type 1 sources are essentially unobscured both optically and in X-rays; intermediate Seyfert types (plotted in purple) show a variety of absorbing columns in between unobscured and obscured sources; and optically obscured sources all have large absorbing columns ($\log \nh{}/{\rm cm}^2 \gtrsim 22.3$). The only X-ray absorbed source with little optical obscuration (NGC~1052) is a LINER that shows broad lines in polarized light \citep{barth1999}, i.e. it is not a ``true type 2'' source \citep[e.g.][]{panessa2002}, but rather one with a peculiar absorbing geometry.

Within the large error of our \av{} estimate, all sources lie above the Galactic ratio for which we assume a range of $\nhavgal = (1.79 \ldots 2.69) \times 10^{21} {\rm cm}^{-2}$ \citep{predehl1995,nowak2012}. We find a large spread in the observed \nhav{} ratio with factors ranging from $\approx$ 1 (NGC~2992) to $\approx 200 \times \nhavgal$ (NGC~1068).

The relation between optical extinction and X-ray absorption was first assessed in a systematic way by \citet{goodrich1994} and \citet{veilleux1997b} and in many other more recent studies using various tracers for the optical extinction. \citet{shi2006} for example find a correlation (with large scatter) between the 9.7 $\mu{}$m silicate strength $\tau_{9.7}$ and the neutral gas column from X-ray observations in the sense that small columns correspond to silicate emission and large columns to silicate absorption. \citet{lyu2014} study the relation between the Balmer Decrement in the {\em Narrow Line Region} and the Silicate strengths in a sample of $\sim$ 100 type 2 AGNs. There is almost no correlation between the two properties which may be related to the fact that the Silicate absorption predominantly arises in the pc-scale torus while the Narrow Line Region is much more extended.

Slightly higher \nh{} values than expected given the optical extinction or reddening have been known since long \citep{mushotzky1982,veroncetty2000} and are also observed in the Galactic Center \citep{porquet2008} (that value is plotted as an asterisk in Fig.~\ref{fig:AVNH}). In the following we discuss possible reasons for deviations from \nhavgal.

\subsection{Increased \nhav{} due to modified dust?}

\citet{maiolino2001} used flux ratios between broad Balmer lines and computed the extinction towards the BLR assuming case B recombination is valid there. They find consistently larger values for \nhav{} than expected from the Galactic diffuse interstellar medium and interpret these offsets as evidence for anomalous dust grains in the dense and extreme environments of galactic nuclei. The largest graphite grains can sustain higher temperatures than smaller silicate grains \citep[e.g.][]{laor1993}. The sublimation ``radius'' is therefore rather a sublimation zone and observations have shown that the innermost dusty region is indeed at smaller radii than expected from models based on Galactic dust properties \citep{kishimoto2007,barvainis1987}. However, this only affects the innermost zone of the dusty region and convincing evidence for altered dust properties in the bulk of the ``torus'' is not available.

\subsection{Increased \nhav{} due to dust-free neutral gas in the BLR?}

An alternative explanation for the observed \nhav{} ratios is that there is considerable absorption by neutral gas clouds inside the dust sublimation zone, such as suggested by \citet{granato1997}. In fact, it is now observationally well established that the neutral absorbing column is variable in many, if not most, AGNs \citep[e.g.][and references therein]{bianchi2012}. The prototypical sources NGC~1365 \citep{risaliti2009a,maiolino2010} and NGC~7582 \citep{bianchi2009} change between Compton-thin and reflection-dominated appearance within hours. Due to the very short timescales, these variations must be caused by clouds in the BLR eclipsing the X-ray source. Further support for dust-free X-ray absorption comes from photoionization models by \citet{netzer2013} which show that the majority of the gas in BLR clouds is expected to be neutral.

In addition to these extreme ``changing-look'' sources, many more sources show changes in their absorbing column over timescales of months and years. Some of these slower changes in absorbing column may also be related to torus clumps passing our line of sight \citep{markowitz2014}, but probably not all of them \citep[e.g.][]{arevalo2014b}. We searched the literature for all kinds of \nh{} variations for the sources in our sample. The range of reported \nh{} values is listed in Tab.~\ref{tab:sample} and indicated in Fig.~\ref{fig:AVNH} by the dashed lines\footnote{For NGC~4388, \citet{elvis2004} reported an uncovering event with an extremely low column of neutral gas ($\log \nh{}/{\rm cm}^2 < 21.9$). We note, however, that their spectral fit also returns a very small photon index ($\Gamma < 1$) which is indicative of a more substantial column than given in that paper. We therefore do not show this value in Fig.~\ref{fig:AVNH}.}. Compared to our ``fiducial value'' of \nh{}, i.e. the consistent analysis based on the spectral fitting in the 0.3-150 keV band by \prepclaudio{}, the variability always changes \nh{} in the expected direction: for low fiducial \nh{} we find variability to values higher than the Galactic ratio and vice versa. We note that due to the variable nature of the gas column in most of the intermediate and obscured sources, the precise ``fiducial'' value of \nh{} is not very important for our conclusions.

It is actually striking to see that {\em all} six intermediate-type sources show variable gas columns (although for two the range is so small that this may also be due to calibration uncertainties). Since not all of the sources have been explored in the same detail, it is quite possible that for many other sources a large part of the gas column is more variable than indicated in the plot.

\subsection{Reliability of our \av{} estimate for the most obscured sources}

Our \av{} estimate relies on the assumption that the 2.3\um{} continuum is dominated by emission from dust at the sublimation radius $r_{\rm sub}$. For type 1 AGNs, this has indeed been shown by observations \citep[e.g.][]{kishimoto2011}. In the most obscured objects, however, the near-infrared emission from $r_{\rm sub}$ is probably obscured and we therefore see hot dust further away from the nuclear engine thanks to a direct sightline through the clumpy torus. In NGC~1068, for example, Speckle imaging and long-baseline infrared interferometry show that the hot dust is located on scales $\approx (3-10) \times r_{\rm sub}$ \citep{weigelt2004,lopezgonzaga2014}. This is also consistent with infrared searches for BLRs in obscured objects which have put large lower limits on \av{} towards the BLR: For five of our type 2 objects, for example, \citet{lutz2002} did not see a broad component of Brackett $\alpha$ (4.05 \um{}) implying $\av{} > 50$. In heavily obscured sources, our \av{} estimate is therefore in fact only a lower limit of the actual \av{} to the inner radius.

Even a twofold increase in the \av{} of these sources would, however, not reduce much the offset these sources show from \nhavgal{}. These sources must have substantial amounts of neutral gas columns {\em within} the dust-free BLR. This is consistent with X-ray observations of multiple thick absorbers along the line of sight \citep[e.g.][for the Circinus~galaxy and NGC~1068, respectively]{arevalo2014b,bauer2015}.

\subsection{Variability in Seyfert type or \av{}?}

Since the X-ray and infrared observations of our sample are not contemporaneous, how likely is it that deviations from \nhavgal{} are caused by variability in the optical obscuration of the AGN? This question is difficult to tackle observationally due to the very long timescales involved. The crossing time for clouds that can obscure a substantial flux of the BLR for example is at least a few years and likely much more \citep{lamassa2015}. Furthermore, in the few, just recently discovered cases where such optical ``changing look'' AGNs have been found by systematic searches in large archives, these changes are attributed to a change in the emission from the central engine rather than obscuring clouds passing by \citep{shappee2014,ruan2015,runnoe2015}. Such major variations in the accretion flow seem to occur in roughly 1 \% of AGNs when comparing observations on a baseline of ten years \citep{macleod2015}. While it is not straightforward to attribute differences in the long-term optical obscuration of an AGN to eclipsing clouds, one can conclude that such changes are rare and therefore unlikely to be the cause for deviations from \nhavgal{} in our sample.

\subsection{Could the obscuration be dominated by large-scale dust lanes?}

Intermediate-type (1.8/1.9) Seyfert galaxies have been found to lie preferentially in edge-on galaxies and this has caused \citet{maiolino1995} to suggest a distinct inner and outer torus, where the latter would be located in the plane of the galaxy and thus be responsible for the obscuration of these sources. This is in line with observations by \citet{prieto2014} who have found that kpc-scale dust lanes can often be traced all the way to the central parsec and may be responsible for some of the obscuration seen in those galaxies. But, on the other hand, based on Hubble Space Telescope ({\em HST}) data of dust structures in a carefully controlled sample of active galaxies, \citet{pogge2002} did not find any evidence for differences in the large scale structure of the various Seyfert subtypes. And on (sub-)parsec scales, mid-IR interferometry of about two dozen objects has shown that the dusty structures in type 1/2 Seyfert galaxies are indistinguishable in terms of size or mid-IR morphology \citep{burtscher2013}.

At any rate, the extinctions \citet{prieto2014} found are only relatively small, in the range \av{} = 3--6 mag. This is sufficient to obscure low luminosity nuclei as the authors note and for some of the less obscured sources in our sample this could be a relevant contribution. It may explain, for example, why the \nhav{} ratio in NGC 2992 is very close to the Galactic ratio: due to its rather high inclination\footnote{Based on 2MASS (K-band) data, the galaxy's axis ratio $b/a$ = 0.43, corresponding to an inclination of about 65 deg out of our line of sight.} its gas and dust column may be dominated by ``standard'' dust in the host galaxy rather than a possibly large dust-free gas column within the BLR.

Large scale dust can therefore add some scatter to our relation at low extinctions, but will not significantly affect the general trend of increasing $N_H$ with $A_V$ over 30 orders of magnitude.



\section{Conclusions}
\label{sec:conclusions}

In this article we re-visit the relation between optical obscuration and X-ray absorption in AGNs. We improve upon previous works by using a novel indicator for the optical obscuration -- the temperature of the hot dust --, consistently estimated X-ray absorbing columns using 0.3--150 keV spectral energy distributions (\prepclaudio{}), by taking into account the variable nature of the neutral gas column \nh{} and by using a consistent classification of Seyfert sub-types. Our findings are summarized here:

\begin{itemize}
	\item All BAT sources without hot dust detections are Compton-thick, but the inverse is not true: some of the most nearby Compton-thick objects show evidence for hot dust (the Circinus~galaxy, NGC~1068 and NGC~7582).
	\item We have collected \oiii/\hb{} fluxes from the literature and determined the Seyfert subtype in a uniform manner using the widely used \citet{winkler1992} scheme. We find that a good correlation exists between Seyfert subtype and \nh{} and that a consistent classification between optical and X-ray appearance can be reached by choosing the thresholds appropriately. For example, if intermediate-type Seyfert galaxies (1.8/1.9) are considered optically unobscured, $\log \nh{}/{\rm cm}^2 > 22.3$ must be required for sources to be X-ray absorbed.
	\item A relation exists between \nh{} and \av{}, where the \nhav{} ratio from the Galactic diffuse interstellar medium sets the floor. Within the errors, all AGNs lie above the relation as seen previously by e.g. \citet{maiolino2001}. By collecting data from the literature, we find that the X-ray absorbing columns in most of the intermediate and obscured AGNs in our sample are variable and their lower values are often nearly in agreement with the expected columns given the optical obscuration and the Galactic ratio. This suggests that the dust properties even in the extreme environment of AGNs can be similar to Galactic. In many sources, deviations from it may simply arise from variable neutral gas clouds within the BLR. This also shows that our ``dust color'' method is a useful tool to derive \av{} in mildly obscured objects.
	\item In the most heavily obscured objects, however, we may underestimate the \av{} towards the dust sublimation radius if the near-IR continuum originates from scales larger than $r_{\rm sub}$ -- an issue that the second generation VLTI instrument GRAVITY \citep{eisenhauer2011} will resolve.
	\item We would encourage X-ray observers to look for significant gas column variability particularly in those sources that show large deviations from the Galactic \nhav{} value, such as in NGC~3081, NGC~4388 or IC~5063.
\end{itemize}

\section*{Acknowledgements}
The authors would like to thank the anonymous referee for comments that helped to improve the paper. LB is supported by a DFG grant within the SPP ``Physics of the interstellar medium''. CR acknowledges financial support from the CONICYT-Chile ``EMBIGGEN'' Anillo (grant ACT1101), from FONDECYT 1141218 and Basal-CATA PFB--06/2007.
This research has made use of the SIMBAD database, operated at CDS, Strasbourg, France and of the NASA/IPAC Extragalactic Database (NED) which is operated by the Jet Propulsion Laboratory, California Institute of Technology, under contract with the National Aeronautics and Space Administration. This research has also made use of NASA's Astrophysics Data System Bibliographic Services and of Astropy, a community-developed core Python package for Astronomy \citep{astropy2013}.

\bibliographystyle{aa}
\bibliography{obscuration.bbl}

\end{document}